\documentclass[12pt]{iopart}
\usepackage{color}
\usepackage{iopams}  
\expandafter\let\csname equation*\endcsname=\relax
\expandafter\let\csname endequation*\endcsname=\relax
\usepackage{amsmath,amssymb,amsthm}
\usepackage{siunitx}
\usepackage{amsmath}	
\usepackage{mathrsfs}
\usepackage{graphicx}
\usepackage{babel}
\usepackage{paralist}
\usepackage{subfigure}
\usepackage{CJK}
\usepackage{lipsum}
\usepackage{soul}

\begin{document}

\title[Absolute frequency measurement of the $^{87}$Sr optical lattice clock at NTSC]{Absolute frequency measurement of the $^{87}$Sr optical lattice clock at NTSC using International Atomic Time}
\author{Xiaotong Lu$^{1}$, Feng Guo$^{1,2}$, Yebing Wang$^{1,2}$, Qinfang Xu$^{1,2}$, Chihua Zhou$^{1,2}$, Jingjing Xia$^{1,2}$, Wenjun Wu$^{1,2}$ and Hong Chang$^{1,2}$}
\address{$\bf{^{1}}$ National Time Service Center, Chinese Academy of Sciences (NTSC), Xi'an 710600, People's Republic of China}
\address{$\bf{^{2}}$ School of Astronomy and Space Science, University of Chinese Academy of Sciences, Beijing 100049, People's Republic of China}
\ead{changhong@ntsc.ac.cn and wuwj@ntsc.ac.cn}
\vspace{10pt}
\begin{indented}
\item[]October 2022
\end{indented}

\begin{abstract}
We report the absolute frequency measurement of the $5s^{2}$ ${}^{1}S_{0}\rightarrow 5s5p$ ${}^{3}P_{0}$ transition in $^{87}$Sr optical lattice clock (Sr1) at National Time Service Center (NTSC). Its systematic frequency shifts are evaluated carefully with a total relative uncertainty of $5.1\times10^{-17}$. The measured absolute frequency is 429 228 004 229 872.91(18) Hz with a relative uncertainty of $4.13\times10^{-16}$, with reference to the ensemble of primary and secondary frequency standards published in the Circular T bulletin by BIPM through a global navigation satellite system (GNSS) link.\\

\end{abstract}


\section{Introduction}
Recent rapid progress of optical clocks based on trapped ions [1-5] and neutral atoms [6–10] has inspired various applications, such as testing possible time variation of fundamental constants [11,12], searching the dark matter [13,14] and applying for relativistic geodesy [15,16]. With estimated systematic uncertainties of optical clocks now routinely lower than those of the most accurate Cs fountains [1-10], the Consultative Committee for Time and Frequency (CCTF) has proposed a roadmap to redefine the second in the International System of Units (SI) [17]. This roadmap requires the absolute frequency measurements of three independent optical clocks to independent Cs primary clocks with fractional uncertainty below $3\times10^{-16}$. In this work, we present the systematic evaluation of the $^{87}$Sr optical lattice clock (OLC) and the absolute frequency measurement using a GNSS link to the ensemble of primary and secondary frequency standards (PSFS) in the BIPM Circular T bulletin.

This paper is organized as follows. Section 2 outlines the optical lattice clock apparatus. Section 3 describes the evaluation of systematic frequency shifts and their uncertainties. Section 4 shows the absolute frequency measurement of the OLC against International Atomic Time (TAI).

\section{Experimental setup}
\begin{figure*}[tbp]
\centering
\includegraphics[width=85mm]{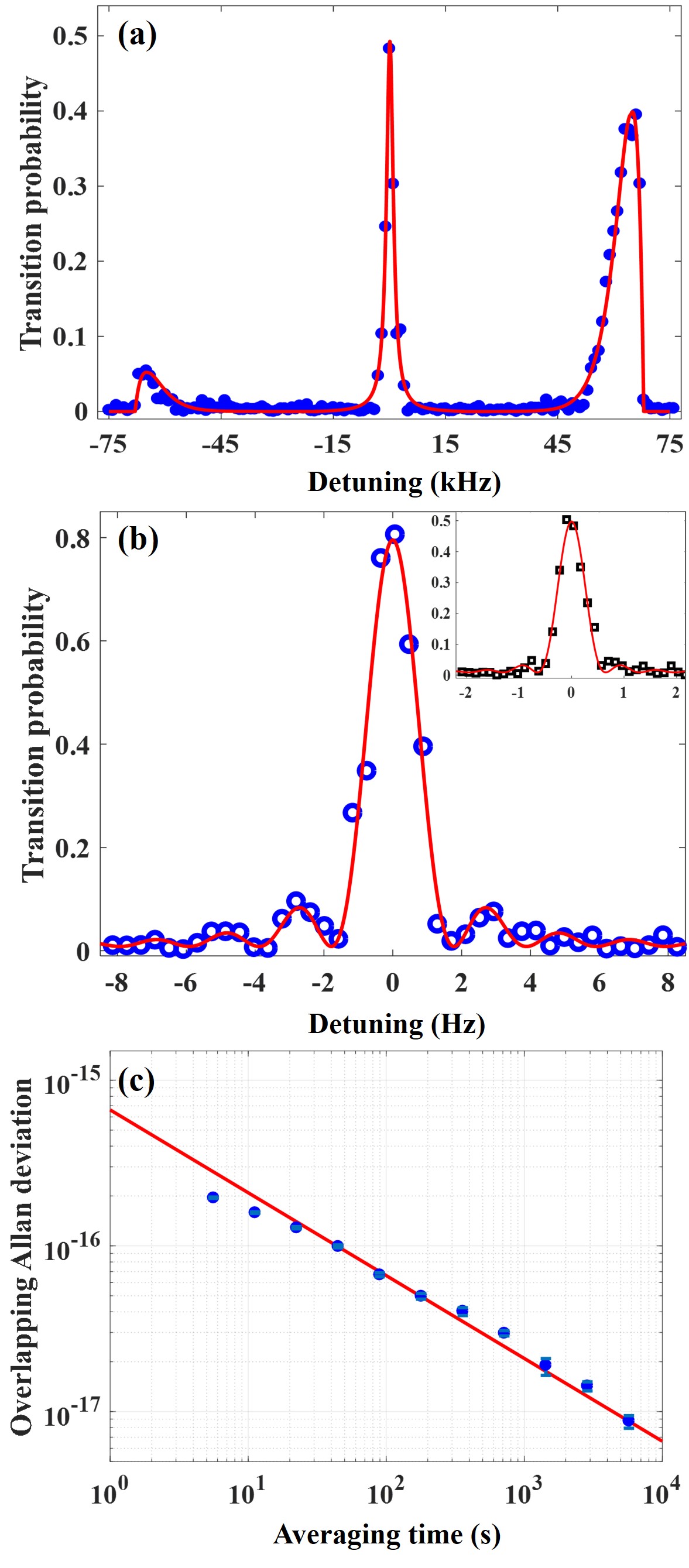}
\caption{\label{fig:exp}(Color online) The clock transition and instability of the $^{87}$Sr OLC. (a) Sideband-resolved spectroscopy. The red line represents the fitting according to reference [22]. (b) High resolution spectrum with a 0.5 s spectroscopy pulse. The inset shows the spectrum with a 1.4 s spectroscopy pulse and the extracted FWHM is 0.59(5) Hz, agreeing with the corresponding Fourier-limited linewidth of 0.57 Hz. The solid lines indicate Rabi fitting. (c) Overlapping Allan deviation of the frequency difference between two interleaved atomic servos at different lattice depths. The comparison data are divided by $\sqrt{2}$ to reflect the interleaved instability. The red fitting line (with white frequency noise model) indicates the interleaved instability of $6.6\times10^{-16}$ $\rm{(\tau/s)^{-0.5}}$.}
\end{figure*}

The setup of the Sr1 at NTSC has been described elsewhere [18-20]. Here, we only give a general overview of the setup. $^{87}$Sr atoms are evaporated from the oven and then collimated by a 0.8 mm diameter pinhole and two-dimensional optical molasses in succession. A Zeeman slower is used to increase the atoms with a velocity below 50 m/s, which is the maximum velocity for capturing atoms by the blue magneto-optical trap (MOT). The blue MOT with a magnetic field gradient of 50 G/cm is operated with light detuned by 40 MHz below the $5s^{2}$ ${}^{1}S_{0}(F=9/2)\rightarrow 5s5p$ ${}^{1}P_{1}(F=11/2)$ transition at 461 nm. Limited by the natural linewidth of the 461 nm transition (about 30.2 MHz), the temperature of atoms is typically $\sim$1 mK after the first stage of cooling by the blue MOT. However, the lattice can capture atoms when the temperature of atoms is tens of $ \mathrm {\mu K} $. Thus, the red MOT, operated with light detuned by $\sim$500 kHz below the $5s^{2}$ ${}^{1}S_{0}(F=9/2)\rightarrow 5s5p$ ${}^{3}P_{1}(F=11/2)$ transition at 689 nm, is used to further reduce atomic temperature to 3 $ \mathrm {\mu K} $. To increase the number of atoms trapped by the red MOT, the light detuned by $\sim$150 kHz below the $5s^{2}$ ${}^{1}S_{0}(F=9/2)\rightarrow 5s5p$ ${}^{3}P_{1}(F=9/2)$ transition at 689 nm is used [18], and the light detuned by $\sim$50 kHz below this transition is used as the spin-polarized light for state preparation. At the end of the red MOT, all light beams except the lattice light are cut off, and about $10^{4}$ atoms are trapped in the lattice. The lattice is formed by overlapping the incident and it's retro-reflected laser beams at the center of the red MOT. The lattice light is generated by an extended cavity diode laser (ECDL), of which the power is amplified by a tapered amplifier (TA). Then, the spin-polarized light with $\sigma^{\pm}$ polarization is used to prepare atoms in $m_{\mathrm F}=\pm9/2$ states. The initial trap depth of the optical lattice is 110 $E_{\rm{R}}$, where $E_{\rm{R}}=h^{2}/2m\lambda^{2}$ is the recoil energy from a lattice photon, $m$ is the mass of the atom, $h$ is the Planck constant, and $\lambda \sim 813.42$ nm is the wavelength of the lattice laser. The hotter atoms are removed by an energy filtering procedure [21], wherein the optical lattice depth is ramped down within 15 ms to 45 $E_{\rm{R}}$, kept for 20 ms and then, ramped back to 110 $E_{\rm{R}}$ within 15 ms. After the energy filtering procedure, approximately 30\% of the atoms remain in the lattice. 

The ultra-stable clock laser corresponds to the $5s^{2}$ ${}^{1}S_{0}\rightarrow 5s5p$ ${}^{3}P_{0}$ transition at 698 nm. Compared to our previous reported clock instability of $5\times10^{-15}$ $\rm{(\tau/s)^{-0.5}}$ [18], we have improved the clock instability by about one order of magnitude by locking the clock laser to a 20 cm ultra-low expansion (ULE) glass cavity. The vacuum pressure of the ULE cavity is $2\times10^{-7}$ Pa maintained by a 100 L/s ion pump. With stabilization  of the power of the cavity transmission light and the temperature of the electro-optic modulator (EOM), the flicker noise floor of the clock laser is determined as $4\times10^{-16}$ by analyzing the in-loop error signal.

The temperature of the ensemble is measured using sideband-resolved spectroscopy shown in figure 1(a), which reveals an axial temperature of $T_{\rm{z}}=1.2$ $\rm{\mu K}$ (indicating that 95\% atoms populate on the ground Bloch band) and a radial temperature of $T_{\rm{r}}=2.3$ $\rm{\mu K}$ [22]. Figure 1(b) shows the high-resolution spectrum with a 0.5 s spectroscopy pulse with a full width at half maximum (FWHM) of 1.6(2) Hz. To explore the limit of coherence in our clock, we scan the clock transition with a 1.4 s interrogation time shown in the inset of figure 1(b). Although the FWHM of the spectrum for 1.4 s clock pulse is 0.59(5) Hz agreeing with the corresponding Fourier-limited linewidth of 0.57 Hz, the transition probability is about 0.5, mainly limited by the interaction between atoms and inhomogeneous excitation due to finite temperature and misalignment angle between the lattice and clock laser beams [22, 23]. In addition, it is difficult for in-loop operation using spectrum with linewidth below 1 Hz. Thus, we set the interrogation time to 0.5 s for our regular operation and the corresponding interleaved instability of $6.6\times10^{-16}$ $\rm{(\tau/s)^{-0.5}}$ is obtained shown in figure 1(c).

\section{Systematic shifts of the $^{87}$Sr clock transition}
There are a number of frequency shifts of the clock transition that must be evaluated when characterizing a lattice clock’s accuracy. Parts of these shifts can be measured by the interleaved self-comparison method, such as the collisional shift, AC Stark shift and DC stark Shift. Other shifts need to combine the theoretical calculation, such as the blackbody radiation (BBR) shift and line pulling shift. A summary of the uncertainty contributions is given in table 1.

\begin{table}[htbp]
\caption{\label{math-tab2}Corrections and uncertainties of Sr1 during the frequency measurement given in fractional frequencies. All values are in units of $10^{-17}$.}
\begin{tabular*}{\textwidth}{@{}l*{15}{@{\extracolsep{0pt plus 12pt}}l}}
\br
Systematic effect&Correction&Uncertainty\\
\mr
BBR chamber&501.9&4.4\\
BBR oven&0.72&0.1\\
Lattice scalar/tensor&5.7&2.2\\
Hyperpolarizability&-1.3&0.5\\
Lattice M1/E2&0&0.4\\
Density&20.32&0.84\\
Tunneling&0&0.6\\
2nd Zeeman&9.39&0.23\\
Background gas collisions&0.51&0.11\\
DC Stark&1.48&0.05\\
Servo error&0.037&0.046\\
Line pulling&0&0.002\\
Probe AC Stark&0.0093&0.0017\\
Total&538.77&5.1\\
\br
\end{tabular*}
\end{table}

\subsection{BBR shift}

The BBR shift is due to the differential Stark shift of the clock states from BBR emitted by the apparatus surrounding the atoms and typically is the largest frequency correction and systematic uncertainty in state-of-the-art optical lattice clocks. The BBR shift of a thermal electric field distribution characterized by a temperature, $T$, can be approximately expressed by [9]
\begin{eqnarray}
\Delta \nu \rm{_{BBR}} = \nu \rm{_{sat}} (\frac{\it{T}}{\it{T}_{0}})^{4}+\nu \rm{_{dyn}} [(\frac{\it {T}}{\it {T}_{0}})^{6}+\vartheta(\frac{\it {T}}{T_{0}})^{8}], \label{eq1}
\end{eqnarray}
where $\nu_{\rm{sat}}$= -2.13023(6) Hz [24], $\nu_{\rm{dyn}}$ = -150.51(43) mHz [25], and $T_{0}$=300 K.

A 10-channel precision temperature scanner (T1001S, XIATECH) and ten calibrated thin-film platinum resistance thermometers (TFPRTs) are used to measure the ambient temperature of the atoms. The total uncertainty of the calibration of TFPRTs is better than 35 mK, where \SI{0}{\degreeCelsius}, \SI{18}{\degreeCelsius}, \SI{22}{\degreeCelsius}, \SI{24}{\degreeCelsius} and \SI{30}{\degreeCelsius} are chosen as the calibration temperatures. Four TFPRTs are put on viewports, covering the top, bottom and lattice viewport. Five TFPRTs are on the stainless-steel chamber and the last one monitors room temperature. Once the maximal and minimal temperature $T_{\rm{max}}$ and $T_{\rm{min}}$ of the system are determined, a representative temperature and an associated uncertainty can be derived, assuming a rectangular probability distribution between $T_{\rm{max}}$ and $T_{\rm{min}}$. During the evaluation, $T_{\rm{max}}$=296.97 K and $T_{\rm{min}}$=294.81 K are obtained. Thus, we use ($T_{\rm{max}}$+$T_{\rm{min}}$)/2=295.89 K as the ambient temperature and ($T_{\rm{max}}$-$T_{\rm{min}}$)/$\sqrt{12}$ =0.624 K as the measurement uncertainty [21,26,27]. Therefore, the BBR shift due to the chamber is evaluated to be $-5.019(44)\times10^{-15}$.
 
The BBR shift from the atomic oven is also estimated. The temperature and solid angle from the oven to MOT are 773(20) K and 0.00009$\rm{\pi}$ sr, respectively. Thus, the BBR shift from the oven is $-7.2(10)\times10^{-18}$. During this evaluation, the heating of the viewport for Zeeman cooling light is stopped. Thus, we will not consider the extra BBR shift caused by this viewport.

\subsection{Density shift}

Previous research has verified that in the spin-polarized Fermion ensemble with a temperature of $\sim$$\rm{\mu K}$, multiply occupied in lattice site will cause density-dependent shift which is dominated by the $p$ wave interaction [28]. Density shift relates to the atomic density, excitation fraction, ensemble temperature and lattice trap depth $U$ [28]. As the ensemble temperature, trap depth and excitation fraction are constant, the density shift is in direct proportion to the number of atoms. Density shift can be measured by the interleaved self-comparison method by modulating the number of atoms of two interleaved loops [29]. In this work, the number of atoms is modulated by a factor of 2 through varying the operation time of the two-dimensional optical molasses between 600 ms and 200 ms, alternatively. When the excitation fraction is 0.418 and $T_{\rm{z}}=1.2$ $\rm{\mu K}$, the density shift is measured at different trap depths shown in figure 2. Using the function of $\alpha_{0}+\alpha U^{1.5}$ to fit the experimental data [6], the coefficients of $\alpha_{0}$ and $\alpha$ are determined as $-0.7(40)\times10^{-6}$ Hz and $-8.94(13)\times10^{-8}$ Hz/$E_{\rm{R}}^{1.5}$, respectively. Thus, the density shift is $-2.032(84)\times10^{-16}$ when the clock regularly runs with $U_{0}$=110 $E_{\rm{R}}$ and $\sim$840 atoms in the lattice.

\begin{figure*}[tbp]
\centering
\includegraphics[width=85mm]{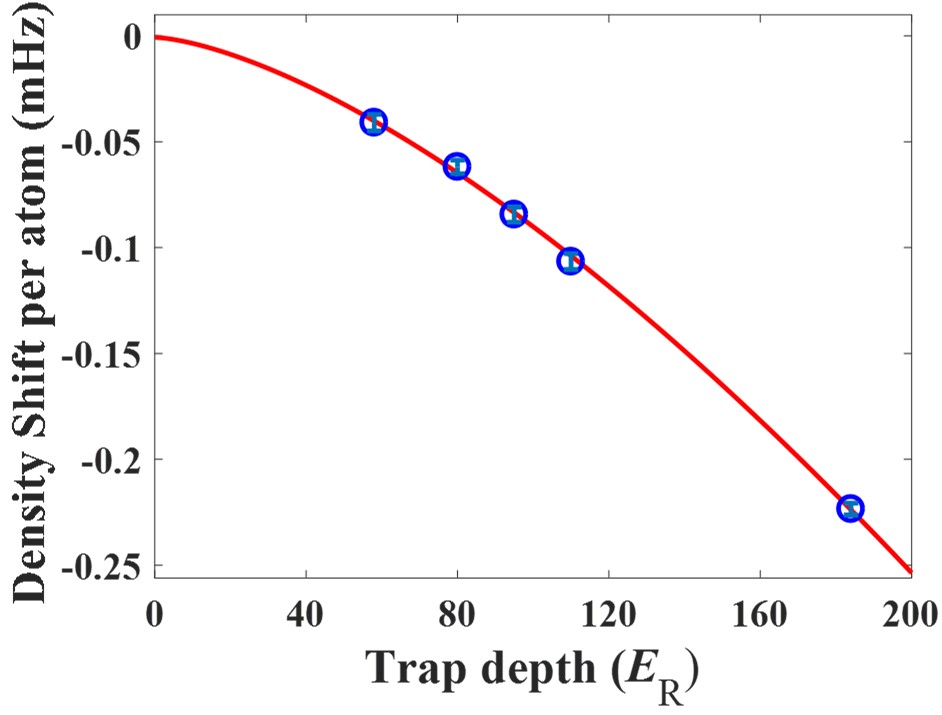}
\caption{\label{fig:exp}(Color online) Density shift evaluated at different trap depths ($U$), scaled to a shift with one atom. The solid red line is a fit to the data with the expected scaling of the shift as $U^{1.5}$. The error bars indicate 1$\sigma$ uncertainty of the corresponding interleaved measurements.}
\end{figure*}

\subsection{Lattice AC Stark shift}

The lattice laser frequency is locked to a 10 cm long ULE cavity with a finesse of about 20000 and measured by a commercial fiber optical frequency comb (OFC, made by IMRA). Before delivering the lattice light to the science chamber, a volume Bragg grating with a bandwidth of 10 GHz is used to reduce the shift caused by the TA amplified spontaneous emission (ASE) [30,31]. The clock laser is co-aligned with the lattice laser. The polarization of the lattice and clock lasers is cleaned by two linear polarizers (LPVIS100-MP2, Thorlabs) and carefully aligned with the vertical bias magnetic field by scanning the $\sigma$ transition spectrum. 

During the total measurements, the lattice laser frequency was set to 368 554 485 MHz for suppressing the first-order AC Stark shift [6]. The residual shift, due to the discrepancy of the tensor shift for different systems [32], is determined using the interleaved self-comparison method. As the density shift coefficient relates to the trap depth, we carefully remove the density shift from the measurement data of the lattice AC Stark shift according to figure 2. By measuring the frequency difference of the clock transition between the lattice trap depths of 110 $E_{\rm{R}}$ and 150 $E_{\rm{R}}$, the scalar/tensor light shift is extrapolated to be $-5.7(22)\times10^{-17}$ for our regular trap depth of 110 $E_{\rm{R}}$, where the uncertainty is inferred from the interleaved instability shown in figure 1(c).

The nonlinear light shift caused by the hyperpolarizability shift and the magnetic dipole and electric quadrupole shifts (M1/E2) have also been evaluated. According to the reference [33], the frequency shift coefficients of hyperpolarizability and M1/E2 terms are 0.46(18) $\rm{\mu Hz}$/$E_{\rm{R}}^2$ and 0(0.31) mHz/$\sqrt{E_{\rm{R}}}$, respectively. In terms of trap depth of 110 $E_{\rm{R}}$ and a mean vibrational state occupation of $\overline{n} = 0.057$, shifts introduced by the hyperpolarizability and M1/E2 term are $1.3(5)\times10^{-17}$ and $0.0(4)\times10^{-17}$, respectively.

\subsection{DC Stark shift}

Stray DC electric fields or patch charges on vacuum viewports can induce DC Stark shift, which was found in optical lattice clocks in 2011 [34]. To remove possible patch charges on vacuum viewports, we use UV light to illuminate each viewport. At the beginning of illuminating the window, the vacuum pressure of the science chamber increases rapidly from $2\times10^{-9}$ Pa to $1\times10^{-7}$ Pa. Keeping illuminating for half an hour, the vacuum pressure returns back to about $5\times10^{-9}$ Pa. To evaluate the residual DC Stark shift using the interleaved self-comparison method, we install 3 pairs of electrodes with a doughnut-like shape to generate the DC electric field. For each direction, we measure the frequency difference between +100 V and 0 V, and -100 V and 0 V, respectively. The total residual DC Stark shift is determined as $-1.48(5)\times10^{-17}$.

\subsection{Zeeman shift}

The average frequency of $m_{\rm{F}}=-9/2 \rightarrow m_{\rm{F}}=-9/2$ and $m_{\rm{F}}=+9/2 \rightarrow m_{\rm{F}}=+9/2$ is used as clock output frequency of the $^{87}$Sr optical lattice clock. To distinguish the two transitions, a large bias magnetic field is applied. The first-order Zeeman shift is canceled out by alternatively locking the clock laser between $m_{\rm{F}}=-9/2$ and $m_{\rm{F}}=+9/2$ components of the clock transition. The frequency gap between $m_{\rm{F}}=-9/2$ and $m_{\rm{F}}=+9/2$ components of the clock transition is 405(5) Hz, where the uncertainty indicates the difference between maximum and minimum frequency gap during 1-month measurements. Thus, the second order Zeeman shift is $-9.39(23)\times10^{-17}$ using the reported coefficient in reference [9].

\subsection{Other shifts and uncertainties}

The site-to-site tunneling of the atoms in the lattice may lead to shift [35]. With the experimental parameters, 99.8\% of the atoms populate the two lowest bands of the lattice (the trap depth is 110 $E_{\rm{R}}$). According to reference [21,36,37], we use the width of the first excited band of the lattice as the upper bound for the effect of tunneling, leading to an uncertainty contribution of $6\times10^{-18}$.

Collision between background gas and atoms will induce frequency shift in OLCs. With the measured vacuum lifetime of 5.9(6) s, this shift is estimated to be $-5.1(11)\times10^{-18}$ using the reported coefficient in reference [38].

The clock transition interrogation is interrupted by detuning the clock laser frequency by 2 MHz from the resonance frequency. This method allows the clock intensity servo to settle before spectroscopy and eliminates all thermally-induced differential path length transients in the acousto-optic modulator (AOM) crystal. The Doppler shift due to path length variation between the clock laser and the atoms trapped in the lattice is suppressed by locking the clock laser phase to the lattice reflector. Thus, the shift caused by the AOM phase chirp and Doppler effect is ignored at the current stage.

We used a second integration of the atomic lock frequency error to compensate for the frequency drift of the clock laser as proposed in reference [39]. The servo error is caused by the residual drift of the clock laser. For the whole measurement time, the mean servo error is $-3.7(46)\times10^{-19}$ by averaging the entire history of atomic servo error signal data. 

The off-resonant excitation of an impurity spin population can produce a systematic line pulling frequency shift. By scanning the clock transition spectrum, the residual total impurity spin population is determined to be less than 4\%. For a 0.5 s $\rm{\pi}$-pulse and a 40.5 Hz splitting between the $|m_{\rm{F}}|=9/2$ and $|m_{\rm{F}}|=7/2$ clock transitions, the maximum off-resonant excitation is: $(0.49/0.82)^{2} \times (\pi/0.5)^{2}/((2\pi×40.5)^{2} + (\pi/0.5))^{2}$ = $2.176\times10^{-4}$ [9]. Combining the upper bound of the atom number in impurity spin states (4\%) with this excitation fraction and the frequency sensitivity of 0.943 Hz [40], an upper bound on the line pulling effect is $2\times10^{-20}$. As the laser is locked to the average frequency of the $\Delta m_{\rm{F}}=0$, which suppresses the shift caused by the line pulling effect. Thus, we make the value of $2\times10^{-20}$ be the uncertainty of the line pulling effect.

The probe Stark shift is in direct proportion to the square of $\Omega_{0}$ ($\Omega_{0}=2 \pi \times 1.3(1)$ Hz for all measurements), where $\Omega_{0}$ indicates the bare states coupling strength between the clock light and atoms. Thus, the probe Stark shift is $-9.3(17)\times10^{-20}$ according to the measured coefficient in our previous publication [41].

\subsection{Systematic shifts summary}

All of the systematic shifts and their uncertainties are summarized in table 1. The total relative systematic shifts correction is $538.77\times10^{-17}$ and the relative uncertainty is $5.1\times10^{-17}$.

\section{Absolute frequency measurement and results}

To accurately determine the absolute frequency of the $^{87}$Sr OLC at NTSC, the H-maser (VCH-1003M, No. 5085) is used as the local transfer oscillator. The 10 MHz output of the H-maser is the reference frequency for OFC and all microwave synthesizers and frequency counters. Both the carrier-envelope offset frequency ($f_{\rm{ceo}}$=60 MHz) and repetition frequency ($f_{\rm{r}}$$\sim$200 MHz) of the OFC are referenced to the H-maser. The 698 nm output of the $^{87}$Sr OLC is delivered to the OFC by a noise-cancelled fiber. $f_{\rm{r}}$ is divided by a factor of 10 and counted by a phase noise analyzer (Microchip 53100A). The beat frequency ($f_{\rm{beat}}$) between the Sr clock output and the corresponding comb tooth is counted by a dual-channel frequency counter (KEYSIGHT 53230A) shown in figure 3(a). The accuracy of these frequency counters was assessed by measuring the 10 MHz maser signal, which is also used as the counter’s external reference, and makes a negligible contribution to the overall uncertainty.

\begin{figure*}[tbp]
\centering
\includegraphics[width=85mm]{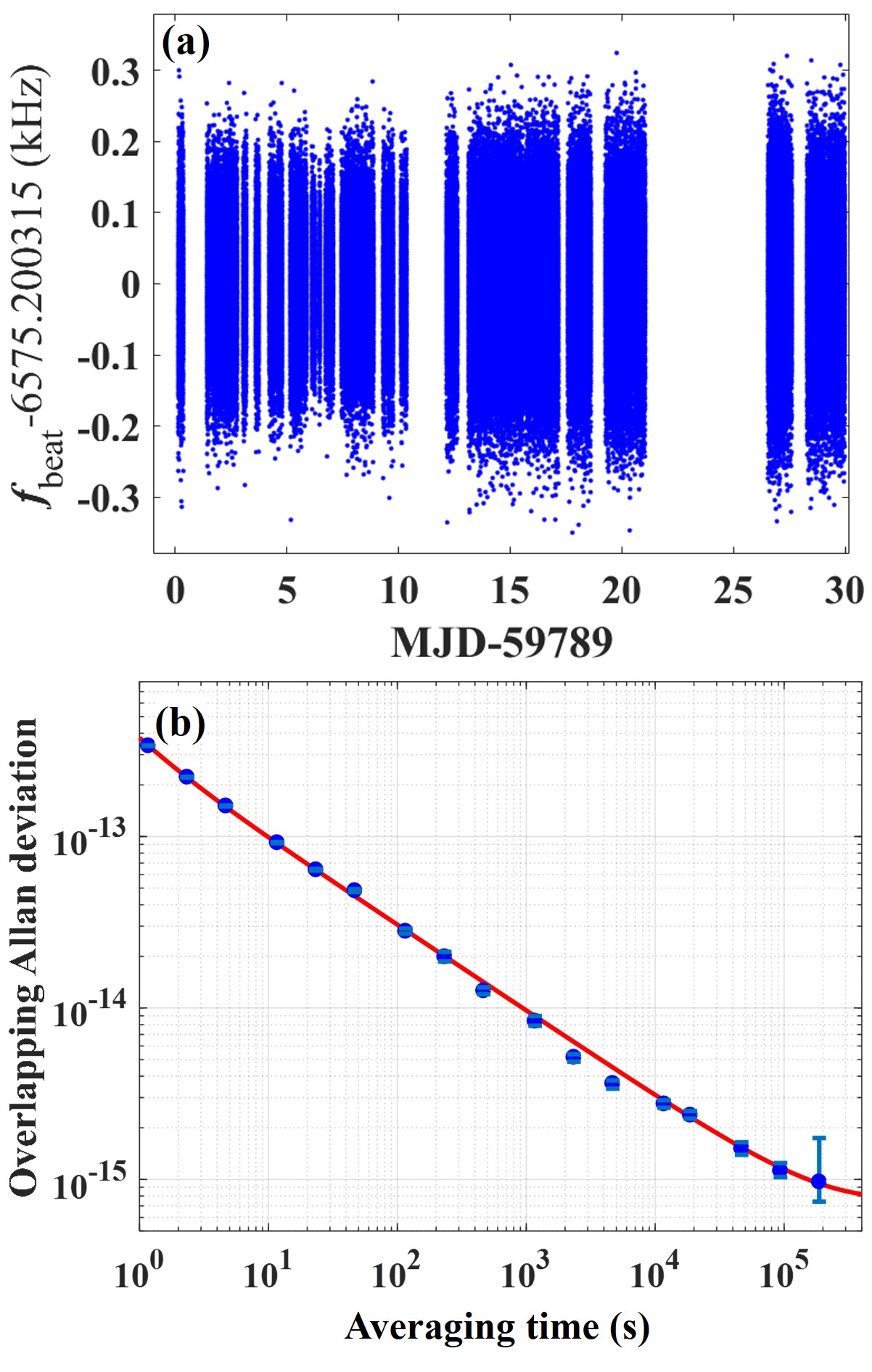}
\caption{\label{fig:exp}(Color online) Optical clock frequency measured with the OFC and the instability of the H-maser. (a) shows the beat frequency $f_{\rm{beat}}$ between the frequency of the $^{87}$Sr OLC transition and the OFC as measured over 30 days. (b) The fractional frequency instability of $f_{\rm{beat}}$ as measured by the overlapping Allan deviation. The Allan deviation points are based on datasets over MJD 59801 to 59810 (with an up-time ratio of 89\%). From this instability data, we extract the stochastic noise model of the H-maser indicated by the red curve.}
\end{figure*}

The H-maser is continually compared to the UTC(NTSC), which is the real-time realization of the coordinated universal time (UTC) by the NTSC, with a local time difference measurement system. UTC(NTSC) has a GNSS link to TAI. According to the monthly Circular T bulletin, the differences between UTC(NTSC), TAI and the best realization of the SI second by PSFS can be determined. Considering inevitable dead times in the measurements, the procedure linking the local optical frequency reference to the SI second can be summarized by the following expression [42,43]
\begin{eqnarray}
\frac{\rm{y_{Sr}}}{\rm{y_{SIs}}} = \frac{\rm{y_{Sr}}(\it{\Delta t}\rm{_1)}}{\rm{y_{HM}}(\it{\Delta t}\rm{_1)}} \times \frac{\rm{y_{HM}}(\it{\Delta t}\rm{_1)}}{\rm{y_{HM}}(\it{\Delta t}\rm{_2)}} \times \frac{\rm{y_{HM}}(\it{\Delta t}\rm{_2)}}{\rm{y_{UTC(NTSC)}}(\it{\Delta t}\rm{_2)}} \nonumber \\ \quad \quad \quad \times \frac{\rm{y_{UTC(NTSC)}}(\it{\Delta t}\rm{_2)}}{\rm{y_{TAI}}(\it{\Delta t}\rm{_2)}}
\times \frac{\rm{y_{TAI}}(\it{\Delta t}\rm{_2)}}{\rm{y_{TAI}}(\it{\Delta t}\rm{_3)}} \times \frac{\rm{y_{TAI}}(\it{\Delta t}\rm{_3)}}{\rm{y_{SIs}}(\it{\Delta t}\rm{_3)}}, \label{eq2}
\end{eqnarray}
where $\rm{y_{Sr}}$, $\rm{y_{HM}}$, $\rm{y_{UTC(NTSC)}}$, $\rm{y_{TAI}}$ and $\rm{y_{SIs}}$ represent the rates of the $^{87}$Sr OLC, H-maser, UTC(NTSC), TAI and the SI second (1 Hz), respectively. $\it{\Delta t}\rm{_i}$ (i = 1, 2, 3) denotes the effective measurement time intervals of the ratios. $\it{\Delta t}\rm{_1}$ is the sum of all effective operation time of Sr1 and OFC measurements; $\it{\Delta t}\rm{_2}$ indicates the minimum range covering $\it{\Delta t}\rm{_1}$, of which the endpoints satisfy the 5-days interval reported in Circular T; $\it{\Delta t}\rm{_3}$ represents the 1-month TAI reporting period.

\begin{table}[htbp]
\caption{\label{math-tab2}The corrections and uncertainties of the calibration of the H-maser frequency. All values are in units of $10^{-16}$.}
\begin{tabular*}{\textwidth}{@{}l*{15}{@{\extracolsep{0pt plus 12pt}}l}}
\br
Ratio&Correction&Uncertainty\\
\mr
$\rm{y_{HM}}(\it{\Delta t}\rm{_1)}/\rm{y_{HM}}(\it{\Delta t}\rm{_2)}$ &-0.083&3.1\\
$\rm{y_{HM}}(\it{\Delta t}\rm{_2)}/\rm{y_{UTC(NTSC)}}(\it{\Delta t}\rm{_2)}$ &1258.8&0.1\\
$\rm{y_{UTC(NTSC)}}(\it{\Delta t}\rm{_2)}/\rm{y_{TAI}}(\it{\Delta t}\rm{_2)}$ &1.54&1.96\\
$\rm{y_{TAI}}(\it{\Delta t}\rm{_2)}/\rm{y_{TAI}}(\it{\Delta t}\rm{_3)}$ &-&-\\
$\rm{y_{TAI}}(\it{\Delta t}\rm{_3)}/\rm{y_{SIs}}(\it{\Delta t}\rm{_3)}$ &-0.9&1.2\\
Total&1259.36&3.9\\
\br
\end{tabular*}
\end{table}

The instability of the comparison between the H-maser and Sr1 is shown in figure 3(b), which can be used to infer the noise model of the H-maser as the instability of Sr1 is smaller than the H-maser by more than two orders of magnitude [25,26,43]. The linear drift rate of the H-maser, considered constant during the one-month measurement, is extracted as $-6.4\times10^{-17}$ $\rm{d^{-1}}$ from the optical frequency measurements. After removing the linear frequency drift, we extracted the four noise contributions of the H-maser with a white phase noise part of $2.21\times10^{-13}$ $(\tau/s)^{-1}$, a white frequency noise part of $3.05\times10^{-13}$ $(\tau/s)^{-0.5}$, a flicker frequency noise part of $6.01\times10^{-16}$ $(\tau/s)^{0}$ and a random walk frequency noise part of $4.49\times10^{-19}$ $(\tau/s)^{0.5}$. Based on this H-maser noise model, we simulated one hundred time series over measurements period $\it{\Delta t}\rm{_1}$ to $\it{\Delta t}\rm{_2}$ using the software of Stable32 [44, 45], and the extrapolation uncertainty of the H-maser frequency is evaluated to be $3.1\times10^{-16}$. The optical frequency measurements are made between MJD 59789.13–59819, which is $\sim$0.13 days shorter than the period covered by $\it{\Delta t}\rm{_2}$. The linear drift of H-maser gives a correction of $-8.3\times10^{-18}$. Combining the dead time and drift of the H-maser, the correction of $\rm{y_{HM}}(\it{\Delta t}\rm{_1)}/\rm{y_{HM}}(\it{\Delta t}\rm{_2)}$ is $-8.3\times10^{-18}$, with an uncertainty of $3.1\times10^{-16}$. The correction of $\rm{y_{HM}}(\it{\Delta t}\rm{_2)}/\rm{y_{UTC(NTSC)}}(\it{\Delta t}\rm{_2)}$ is calculated to be $1.2588(1)\times10^{-13}$ from local DMTD measurements. As $\it{\Delta t}\rm{_3}=\it{\Delta t}\rm{_2}$, it is unnecessary to evaluate the correction of $\rm{y_{TAI}}(\it{\Delta t}\rm{_2)}/\rm{y_{TAI}}(\it{\Delta t}\rm{_3)}$. With the data from Circular T 416, the corrections of $\rm{y_{UTC(NTSC)}}(\it{\Delta t}\rm{_2)}/\rm{y_{TAI}}(\it{\Delta t}\rm{_2)}$ is calculated to be $1.54\times10^{-16}$, of which the uncertainty can be determined by $(5/30)^{0.9} \times \sqrt{2} \times 0.3/86400/5$ = $1.96\times10^{-16}$ [26]. The correction of $\rm{y_{TAI}}(\it{\Delta t}\rm{_3)}/\rm{y_{SIs}}(\it{\Delta t}\rm{_3)}$ is $-0.9(12)\times10^{-16}$. The calibration of the H-maser has a correction factor of $1.25936\times10^{-13}$ with an uncertainty of $3.9\times10^{-16}$, as summarized in table 2. 

\begin{table}[htbp]
\caption{\label{math-tab2}The corrections and uncertainties of the absolute frequency measurement. All values are in units of $10^{-16}$.}
\begin{tabular*}{\textwidth}{@{}l*{15}{@{\extracolsep{0pt plus 12pt}}l}}
\br
Ratio&Correction&Uncertainty\\
\mr
Sr1 systematic&53.877&0.51\\
Statistical&0&1.1\\
Relativistic redshift&-523.7&0.6\\
H-maser calibration&1263.03&3.9\\
Total&793.207&4.13\\
\br
\end{tabular*}
\end{table}

The standard error of the weighted mean of all measurements shown in figure 3(a) is $1.1\times10^{-16}$, which has been scaled by the root of the reduced chi-square $\chi_{red}^2$ ($\chi_{red}^2=2.8$) and used to be the statistical uncertainty of the optical frequency measurement. The orthometric height of the lattice is 480.3(5) m by tracing to a leveling benchmark inside the building on our campus. This leveling benchmark is referenced to the China 1985 national height datum [46]. Benefiting from the one-month-long measurement with an up-time ratio of 57.1\%, the time-variable gravity potential component is averaged out and considerable to be negligible [47]. With the method in reference [48], the fractional relativistic redshift is $523.7(6)\times10^{-16}$. Thus, the total calibration of the whole measurement system is $793.207\times10^{-16}$ with an uncertainty of $4.13\times10^{-16}$. The uncertainties of the absolute frequency measurement are summarized in table 3. The absolute frequency of Sr1 is 429 228 004 229 872.91(18) Hz, which is consistent with the 2021 CIPM recommended value of the $^{87}$Sr neutral atom optical frequency of 429 228 004 229 872.99(7) Hz.

\section{Conclusion}

In summary, we have realized an $^{87}$Sr optical lattice clock with an estimated systematic uncertainty of $5.1\times10^{-17}$ and determined its absolute frequency to be 429 228 004 229 872.91(18) Hz by tracing its frequency to the fountain ensemble in Circular T through a GNSS link with an overall uncertainty of $4.13\times10^{-16}$. Future works will focus on improving the effective operation time of the optical lattice clock, reducing the systematic uncertainty and comparing against other optical clocks with a comparison uncertainty below the $1\times10^{-17}$ level.

\section{Acknowledgments}
This work is supported by the Strategic Priority Research Program of the Chinese Academy of Sciences (Grant No. XDB35010202) and the National Natural Science Foundation of China (No. 12203057).

\section{Data availability}
The data that support the findings of this study are available upon reasonable request from the authors.

\section{References}

\numrefs{1}

\item Chou C W, Hume D B, Koelemeij J C J, Wineland D J and Rosenband T 2010 {\it Phys. Rev. Lett.} {\bf 104} 070802
\item Huntemann N, Sanner C, Lipphardt B, Tamm C and Peik E 2016 {\it Phys. Rev. Lett.} {\bf 116} 063001
\item Brewer S M, Chen J S, Hankin A M, Clements E R, Chou C W, Wineland D J, Hume D B and Leibrandt D R 2019 {\it Phys. Rev. Lett.} {\bf 123} 033201
\item Cui K F, Chao S J, Sun C L, Wang S M, Zhang P, Wei Y F, Yuan J B, Cao J, Shu H L and Huang X R 2022 {\it Eur. Phys. J. D} {\bf 76} 140
\item Huang Y, Zhang B L, Zeng M Y, HaoY M, Ma Z X, Zhang H Q, Guan H, Chen Z, Wang M and Gao K L 2022 {\it Phys. Rev. Applied} {\bf 17} 034041
\item Nicholson T L et al 2015 {\it Nat. Commun.} {\bf 6} 6896
\item Ushijima I, Takamoto M, Das M, Ohkubo T and Katori H 2015 {\it Nat. Photonics} {\bf 9} 185-9
\item McGrew W F et al 2018 {\it Nature} {\bf 564} 87-90
\item Bothwell T, Kedar D, Oelker E, Robinson J M, Bromley S L, Tew W L, Ye J and Kennedy C J 2019 {\it Metrologia} {\bf 56} 065004
\item Lu B K, Sun Z, Yang T, Lin Y G, Wang Q, Li Y, Meng F, Lin B K, Li T C and Fang Z J 2022 {\it Chin. Phys. Lett.} {\bf 39} 080601
\item Huntemann N, Lipphardt B, Tamm C, Gerginov V, Weyers S and Peik E 2014 {\it Phys. Rev. Lett.} {\bf 113} 210802
\item Godun R M et al 2014 {\it Phys. Rev. Lett.} {\bf 113} 210801
\item Derevianko A and Pospelov M 2014 {\it Nat. Phys.} {\bf 10} 933-6
\item Roberts B M et al 2020 {\it New J. Phys.} {\bf 22} 093010
\item Lisdat C et al et al 2016 {\it Nat. Commun.} {\bf 7} 12443
\item Grotti J et al 2018 Nat. {\it Phys.} {\bf 14} 437–41
\item Riehle F, Gill P, Arias F and L. Robertsson, 2018 {\it Metrologia} {\bf 55} 188
\item Wang Y B, Yin M J, Ren J, Xu Q F, Lu B Q, Han J X, Guo Y and Chang H 2018 {\it Chin. Phys. B} {\bf 27} 023701
\item Wang YB, Lu B Q, Kong D H and Chang H 2018 {\it Appl. Sci.} {\bf 8} 2194
\item Lu X T, Zhou CH,  Li T, Wang YB and Chang H 2020 {\it Appl. Phys. Lett.} {\bf 117} 231101
\item Falke S et al 2014 {\it New J. Phys.} {\bf 16} 073023
\item Blatt S, Thomsen J W, Campbell G K, Ludlow A D, Swallows M D, Martin M J, Boyd M M and Ye J 2009 {\it Phys. Rev. A} {\bf 80} 052703
\item Bishof M, Martin M J, Swallows M D, Benko C, Lin Y, Quemener G, Rey A M and Ye J 2011 {\it Phys. Rev. A} {\bf 84} 052716
\item Middelmann T, Falke S, Lisdat C and Sterr U 2012 {\it Phys. Rev. Lett.} {\bf 109} 263004
\item Schwarz R, D$\rm{\ddot{o}}$rscher S, Al-Masoudi A, Benkler E, Legero T, Sterr U, Weyers S, Rahm J, Lipphardt B and Lisdat C 2020 {\it Phys. Rev. Research} {\bf 2} 033242
\item Hobson R, Bowden W, Vianello A, Silva A, Baynham C F A , Margolis H S, Baird P E G, Gill P and Hill I R 2020 {\it Metrologia} {\bf 57} 065026
\item Pizzocaro M, Thoumany P, Rauf B J, Bregolin F, Milani G, Clivati C, Costanzo G A, Levi F and Calonico D 2017 {\it Metrologia} {\bf 54} 102-112
\item Rey A M, Gorshkov A V, Kraus C V, Martina M J, Bishof M, Swallows M D, Zhang X, Benkoa C, Ye J, Lemke N D and Ludlow A D 2014 {\it Ann. Phys.} {\bf 340} 311-351
\item Nicholson T L, Martin M J, Williams J R, Bloom B J, Bishof M, Swallows M D, Campbell S L and Ye J 2012 {\it Phys. Rev. Lett.} {\bf 109} 230801
\item Fasano R J, Chen Y J, McGrew W F, Brand W J, Fox R W and Ludlow A D 2021 {\it Phys. Rev. Applied} {\bf 15} 044016
\item Takamoto M, Ushijma, Ohmae N, Yahagi T, Kokado K, Shinkai H and Katori H 2020 {\it Nat. Photon.} {\bf 14} 411-415 
\item Shi C, Robyr J L, Eismann U, Zawada M, Lorini L, Targat R Le and Lodewyck J 2015 {\it Phys. Rev. A.} {\bf 92} 012516
\item Westergaard P G, Lodewyck J, Lorini L, Lecallier A, Burt E A, Zawada M, Millo J and Lemonde P 2011 {\it Phys. Rev. Lett.} {\bf 106} 210801
\item Lodewyck J, Zawada M, Lorini L, Gurov M and Lemonde P 2012 {\it IEEE Trans. Ultrason. Ferroelectr. Freq. Control} {\bf 59} 411-5
\item Lemonde P and Wolf P 2005 {\it Phys. Rev. A} {\bf 72} 033409
\item Yin M J, Lu X T, Li T, Xia J J, Wang T, Zhang X F and Chang H 2022 {\it Phys. Rev. Lett.} {\bf 128} 073603
\item Bloch I, Dalibard J and Zwerger W 2008 {\it Rev. Mod. Phys.} {\bf 80} 885-964
\item Alves B X R, Foucault Y, Vallet G and Lodewyck J 2019 Joint Conf. of the IEEE Int. Frequency Control Symp. and European Frequency and Time Forum pp 1-2
\item Peik E, Schneider T and Tamm C 2006 {\it J. Phys. B: At. Mol. Opt. Phys.} {\bf 39} 145
\item Al-Masoudi A, D$\rm{\ddot{o}}$rscher S, Hafner S, Sterr U and Lisdat C 2015 {\it Phys. Rev. A} {\bf 92} 063814 
\item Xu Q F, Lu X T, Xia J J, Wang Y B and Chang H 2021 {\it Appl. Phys. Lett.} {\bf 119} 101105
\item Hachisu H, Petit G, Nakagawa F, Hanado Y and Ido T 2017 {\it Opt. Express} {\bf 25} 8511
\item Lin Y G, Wang Q, Meng F, Cao S Y, Wang Y Z, Li Y, Sun Z, Lu B K, Yang T, Lin B K, Zhang A M,Fang F and Fang Z J 2021 {\it Metrologia} {\bf 58} 035010 
\item Yu D-H, Weiss M and Parker T E 2007 {\it Metrologia} {\bf 44} 91-96
\item Hachisu H and Ido T 2015 {\it Jpn. J. Appl. Phys.} {\bf 54} 112401
\item Li J C, Chu Y H and Xu X Y 2017 Determination of vertical datum offset between the regional and the global height datum {\it Acta Geod. Cartogr. Sinica} {\bf 46} 1262
\item Voigt C, Denker H and Timmen L 2016 {\it Metrologia} {\bf 53} 1365-83 
\item Denker H, Timmen L, Voigt C, Weyers S, Peik E, Margolis H S, Delva P, Wolf P and Petit G 2018 {\it J. Geod.} {\bf 92} 487-516 

\endnumrefs

\end{document}